\documentclass[graybox]{svmult}

\usepackage{type1cm}        
\usepackage{makeidx}         
\usepackage{graphicx}        
\usepackage{multicol}        
\usepackage[bottom]{footmisc}

\usepackage{newtxtext}       %
\usepackage[varvw]{newtxmath}       

\makeindex             

\begin{document}

\title*{Exploring Cybercriminal Activities, Behaviors and Profiles}
\author{Maria Bada and Jason R.C. Nurse}
\institute{Maria Bada \at School of Biological and Behavioural Sciences, Queen Mary University of London, E1 4NS, London, UK.  \email{m.bada@qmul.ac.uk}\newline
*Corresponding author.
\and Jason R.C. Nurse \at School of Computing, 
University of Kent, Canterbury, 
CT2 7NF, Kent, UK. \email{j.r.c.nurse@kent.ac.uk} 
}
%
%
\maketitle

\abstract*{While modern society benefits from a range of technological advancements, it also is exposed to an ever-increasing set of cybersecurity threats. These affect all areas of life including business, government and individuals. To complement technology solutions to this problem, it is crucial to understand more about cybercriminal perpetrators themselves, their use of technology, psychological aspects, and profiles. This is a topic that has received little socio-technical research emphasis in the technology community, has few concrete research findings, and is thus a prime area for development. The aim of this article is to explore cybercriminal activities and behavior from a psychology and human aspects perspective, through a series of notable case studies. We examine motivations, psychological and other interdisciplinary concepts as they may impact/influence cybercriminal activities. We expect this paper to be of value and particularly insightful for those studying technology, psychology, and criminology, with a focus on cyber security and cybercrime.}

\abstract{While modern society benefits from a range of technological advancements, it also is exposed to an ever-increasing set of cybersecurity threats. These affect all areas of life including business, government and individuals. To complement technology solutions to this problem, it is crucial to understand more about cybercriminal perpetrators themselves, their use of technology, psychological aspects, and profiles. This is a topic that has received little socio-technical research emphasis in the technology community, has few concrete research findings, and is thus a prime area for development. The aim of this article is to explore cybercriminal activities and behavior from a psychology and human aspects perspective, through a series of notable case studies. We examine motivations, psychological and other interdisciplinary concepts as they may impact/influence cyber-criminal activities. We expect this paper to be of value and particularly insightful for those studying technology, psychology, and criminology, with a focus on cybersecurity and cybercrime.}

\textbf{Keywords}: Cybersecurity, cyber psychology, cognition, human aspects, cybercrime, cybercriminal, online offender, behavior. 

\section{Introduction}
\label{sec:1}
Cybercrime has grown substantially in the last 18 months, and has impacted businesses, members of the public, and governments alike. While the trajectory of cyber-attacks has been on the rise for a number of years, the increased digitization that has emerged a result of COVID-19 (SARS-CoV-2), the stress and uncertainty caused in the population because of the pandemic, and the general challenges to securing remote workforces, has led to significant issues of online crime~\cite{kar87308,lallie2021cyber}. One study has reported that cybercrime has increased 600\% due to COVID-19 pandemic \cite{purplesec2021} and in some countries (e.g., the UK) this rise has led to record numbers of attacks faced by the society~\cite{sabbagh2021covid}. International and regional policing organizations (e.g., Interpol and Europol) have thus warned businesses and individuals about these attacks, and released guidance on staying safe and boosting threat response and cyber hygiene. 

To understand the nature of cybercrime, it is imperative to examine the threat ac-tors or offenders behind the crimes, what motivates them, their behaviors and profiles. This area of research has often been referred to as cybercriminal (or online offender) understanding or profiling, and tends to mirror the offline, and more traditional action of criminal profiling~\cite{turvey2011criminal}. In this regard, and extending upon prior works \cite{badanurse2021cy,jahankhani2012examination,warikoo2014proposed}, we consider cybercriminal profiling/understanding to generally be an educated attempt to define information about the person who committed a cybercrime, which considers their characteristics, patterns or other factors of uniqueness. 

While there has been an increasing amount of research in the cybercriminal space, this topic that has received little socio-technical research emphasis in the technology community, has few concrete research findings, and is thus a prime area for development. Bada \& Nurse~\cite{badanurse2021cy} summarized the outstanding challenges with specific mention of the need to explore the actions and personality traits apparent from certain online criminal behaviors; a factor also driven by the lack of studies drawing on actual data linked to behavioral profiles.

The aim of this article therefore is to investigate cybercriminal activities and behavior from a socio-technical (psychology and human aspects) perspective, through reflecting on the state of the art as well as a series of notable cybercriminal case studies. This work considers the motivations of online offenders, and psychological and other interdisciplinary concepts as they may impact/influence cybercriminal actions. The remainder of this contribution is as follows. Section~\ref{sec:2} reflects on the threat of cybercrime more broadly, outlines the main types of attack and the traditional threat actors commonly discussed in research and practice. Section~\ref{sec:3} examines several cybercriminal cases in detail drawing on cyberpsychology, cognition, human aspects and cybersecurity research, to identify characteristics and profiles of offenders. Finally, Section~\ref{sec:4} concludes this article and highlights key aspects when exploring cybercriminal activities, behaviors and profiles.

\section{The Threat of Cybercrime: Actions and Actors}
\label{sec:2}
Cybercrime is often used in the media and in research to refer to a range of crimes conducted in the online (or cyber) space. The reality, however, is that these crimes are extremely varied. The UK’s Crown Prosecution Service (CPS) deconstructs cybercrimes in two primary types: Cyber-dependent crimes and Cyber-enabled crimes. Cyber-dependent crimes are: ``crimes that can be committed only through the use of Information and Communications Technology (`ICT') devices, where the devices are both the tool for committing the crime, and the target of the crime (e.g. developing and propagating malware for financial gain, hacking to steal, damage, distort or destroy data and/or network or activity)''~\cite{cps2019cyb}. These include hacking, causing disruption due to malware, and the use of botnets for service disruption. Alternately, Cyber-enabled crimes are: ``traditional crimes which can be increased in scale or reach by the use of computers, computer networks or other forms of ICT''~\cite{cps2019cyb}. Examples of these crimes include online fraud, data theft, cyber harassment, and child sexual offences. 

The characterization above has evolved since early work on cybercrime (e.g., \cite{gordon2006definition}) but there are still various similarities, particularly the focus on technology versus other aspects (Gordon \& Ford for instance, refer to a continuum with technology crime on one side and people crime on the other~\cite{gordon2006definition}). Other research overlooks high-level categorizations and concentrates on the specific actions/crimes. Relevant examples include Stabek et al.~\cite{stabek2010seven} who examine specific scam types, Nurse~\cite{nursecybercrime} that explores crimes against individuals, and Chiew et al.~\cite{chiew2018survey} who assess the nature (types, vectors, approaches) of phishing attacks.

Behind criminal actions (be they referred to as crimes or cyber-attacks) are perpetrators who are responsible for planning, orchestration or execution. Initial characterizations of these individuals centred on high-level groupings, such as script kiddies, hackers, fraudsters, insider threats, hacktivists, and nation states. In that context, script kiddies were typically viewed as the lowest skilled and resourced, while nation states were at the other end of the spectrum. 

Today, online offenders and attack perpetrators share some similarities with the groupings above but their profiles are also often much more nuanced. For instance, research has examined the psyche of cybercriminals~\cite{barnor2020rationalizing,kirwan2013cybercrime,rogers2011psyche} and the theories behind why cybercrime occurs~\cite{palmieri2021personality,stalans2018explaining}, and other work has investigated attackers in depth---be it on the presence of the hacktivist group Anonymous online~\cite{jones2020behind} or nation state Advanced Persistent Threats (APTs)~\cite{nikkel2021apts}. Considering the psychology of perpetrators themselves, online criminal behavior has been related to psychopathy and other antisocial behaviors~\cite{seigfried2017computer}, persons high on Machiavellianism (one of the three Dark Triad personality traits) have been shown as more likely to engage in criminal behavior~\cite{selzer2021saint}, and we have found relationships cited between cybercriminal actions and conditions such as autism~\cite{lim2021revisiting}. These all point to the importance of exploring perpetrators as a part of understanding cybercrime. 

Theories of crime developed by the field of cyberpsychology such as the online disinhibition effect~\cite{Suler2004} can also be considered relevant to understanding why an individual may engage in online criminal acts, however, its usefulness depends on the type of cybercrime considered.
Neutralizations~\cite{sykes1957techniques} from offenders offering explanations for crimes that they would normally consider to be morally unacceptable are common in different types of crime including cybercrime. Such excuses can include denying responsibility for their actions or denial of injury to the victim. In summary, the reality is that developing a better understanding of the persons behind cybercrimes is key for research and practice.

\section{Cybercriminal Case Studies}
\label{sec:3}

\subsection{Overview and Method of Analysis}
In this study our method of analysis drawns on the different factors and abilities described in models such as the Deductive Cybercriminal Profile Model~\cite{Nykodym2005} and the Theoretical Model of Profiling a Hacker~\cite{Lickiewicz}. These models guide the collection of information required in order to create a holistic profile. In general, they propose that in order to form a psychological profile of an offender, different factors need to be considered: a) biological factors and the external environment which influences an individual; b) intelligence; c) personality; d) social abilities; and e) technical abilities. The theoretical model of profiling a hacker~\cite{Lickiewicz} also includes factors such as: f) motivation for offending; g) the method of the attack; and h) the effectiveness of the attack.

Below we will present cases of persons identified in the literature (at one point or another) as real cyber offenders, and describe their characteristics, traits, motivations and behaviors. This approach will also allow for a reflection on the similarities and differences among the different cases. When analysing the cases, theories of the Dark Triad/Tetrad~\cite{DarkTriad}, the HEXACO model of personality~\cite{HEXACO} and theories of crime will be utilised as well. Readers should note that we intentionally do not directly name the persons that we present given the sensitivity of the topic. Moreover, we present point in time analyses based on literature and existing reports. This is worth noting because people change (e.g., some once-famous hackers are now well-respected security professionals), and secondly, we rely on reports for our reflection (thus, rely on the accuracy of the reports we draw on).

\subsection{Case 1} 
Case 1 was known as the first cybercriminal in the US, releasing the first ‘worm’ on the internet in 1988 whilst attending Cornell University~\cite{FBInews}. Utilising the Unix Sendmail program, he reportedly altered it to replicate itself, and it caused computers to crash (with as many as 6,000 computers impacted). 

\textit{Skills}: 
Case 1 studied computer science and graduated from Harvard. At Harvard, Case 1 was reportedly known for his technological skills, but also his social skills ~\cite{FBInews}. After graduating, he continued his studies at Cornell; he later developed a malicious program which was released via a hacked MIT computer~\cite{FBInews}.

\textit{Characteristics and Psychological Traits}:
Case 1's father was an early innovator at a technology lab so he grew up immersed in computers~\cite{FBInews}. Case 1 reportedly was the type of student who found homework boring and therefore focused his energy in programming; he also preferred to work alone~\cite{TheWashingtonPost}. This rather agrees with findings indicating that personality traits such as introversion are associated to online criminal behaviour ~\cite{seigfried2017computer}. 

\textit{Motivation}: 
According to reports, Case 1  claimed that his actions did not have malicious intent but rather his aim was to point out the safety issues and vulnerabilities of systems~\cite{OKTA}. The worm did not damage or destroy any files, but it slowed down University functions causing substantial economic losses~\cite{FBInews}. The network community tried several techniques in order to understand the worm and to remove it from their systems. Some of the affected institutions disconnected their computers while others had to reset their systems. Case 1, however, was not imprisoned but he was sentenced to probation for three years and also community service~\cite{OKTA}. 

\subsection{Case 2} 
Case 2 was a teen hacker, a computer programmer and the founder of an non-profit organisation that publishes leaks. As covered by~\cite{TheGuardian}, during his studies in Australia he lived in a student house where he spent much of his time dreaming of setting up a new way to disseminate classified information. By 1991 Case 2 was reportedly one of the most accomplished hackers in Australia~\cite{TheGuardian}.

\textit{Skills}: 
He was characterised by high, analytical intelligence. In 1991, he reportedly formed a hacking group called the International Subversives~\cite{AtlanticCouncil}. During this time, he hacked into Military Institutions, such as MILNET, the US military's secret defence data network, and Universities~\cite{TheGuardian}. According to reports, his father had a highly logical intellect which Case 2 is said to have inherited from him~\cite{TheRegister}.

\textit{Characteristics and Psychological Traits}: 
As a student, articles (e.g.,~\cite{TheGuardian}) note that Case 2 was not interested much in the school system. In terms of his personality, resources state that he lacked social skills, had a dry sense of humour, and at times also often forgot basic hygiene behaviors~\cite{TheGuardian}. 

Case 2 reportedly disregarded those he disapproved of, he could easily get angry, and had instant mood changes~\cite{TheGuardian}. Eysenck’s Theory of Crime proposes that personality traits such as Psychoticism (being anti-social, aggressive and uncaring), Extraversion (seeking sensation) and Neuroticism (being unstable in behavioral patterns) indicate a personality susceptible to criminal behavior~\cite{Eysenck1964}. However, in Case 2 we may also see a similar pattern as in Case 1, a sense of superiority seen in narcissistic personalities~\cite{DarkTriad}. 

\textit{Motivation}:
In terms of the motive behind Case 2, according to the prosecution during his trial at the Victoria County Court in Melbourne, it was ``simply an arrogance and a desire to show of his computer skills''~\cite{TheGuardian}. Case 2 pleaded guilty to 24 counts of hacking~\cite{TheGuardian}.

\subsection{Case 3} 
Case 3 was a known ex-member of the group Anonymous. This group is referred to as hacktivists, who utilise sometimes criminal acts as a way to pursue particular motives. Case 3 was found to be a member when he gave his identity willingly to the police during an attempt to pursue justice in a rape case~\cite{Anonumous}. The back story involves a football team in the US that was accused of raping a 16-year-old girl, but were not prosecuted, despite evidence (see~\cite{Kushner}). This led to Anonymous' hacking of the football website and email of someone affiliated with the team, revealing indecent images of young women. Case 3 was noted to be one of the main activists behind these events~\cite{Kushner}.

\textit{Skills}: 
While Case 3 reportedly dropped out of school, he showed a keen interest in computers/technology, teaching himself how to code for instance~\cite{Kushner}. 

\textit{Characteristics and Psychological Traits}: 
Case 3 was reported being shy and a frequent target of bullying at school, experiencing violent episodes during adulthood~\cite{Kushner}. Research~\cite{Wolke2015} has suggested that  bullying is linked to altered cognitive responses to stressful and threatening situations. Further, \cite{ORiordan} noted that the presence of school problems during adolescence may contribute to criminal behavior. Case 3 was reportedly unstable during his teenage years, he formed a gang to bully the bullies, had drinking issues and spent some time homeless~\cite{Esquire}.
These behaviors could potentially indicate personality traits such as neuroticism and psychoticism, as defined by Eysenck’s theory~\cite{Eysenck1964}. In addition, as the Five Factor Model~\cite{Costa2002} and the HEXACO Model~\cite{HEXACO} describe, an individual low in agreeableness may tend to be critical, hostile and aggressive. In this case these traits may be portrayed by being critical to others and speaking of injustice.

\textit{Motivation}: 
Case 3 claimed his motives were for justice, defending the victims being targeted. He spoke of a few cases of hacking he conducted under the signature Anonymous mask~\cite{Kushner}. He claimed he would target bullies; those who also used technology to harm others. Reflecting on this case, there is again a possible implication that he was better suited than law enforcement to manage such a situation. This self-justification of labelled criminal acts potentially suggests narcissistic personality traits~\cite{DarkTriad}. 

It is likely that this individual found a sense of power through hacking, something he may have not had as a child when he himself was the victim. Reports~\cite{Esquire} note that Case 3 optimised the overall Anonymous group persona, hiding his face, creating a false name, and posting videos online with distortions to protect his identity. It is such a persona that can facilitate such behavior online~\cite{Suler2004}. 

\subsection{Case 4} 
Case 4 was a hacktivist reported to be responsible for hacking into a large number of government computer systems, such as the FBI, stealing large amounts of data~\cite{TheGuardian2017}. 

\textit{Skills}: Activism and hacking were a noteworthy theme in Case 4's life. According to resources, by the age of 8, he had enough skills to rewrite the computer code of applications~\cite{Computerweekly}.  Case 4 and his sibling, enjoyed playing video games and this led them into finding techniques to cheat the technology so that they would always win~\cite{TheTelegraph}. Early on during his education he appears to have become bored, and in lower school he was assigned a dedicated tutor because, as he stated, ``there was nothing left to teach me on the curriculum'' ~\cite{Computerweekly}. Case 4 studied computer science at A-level and at university. As he stated, ``One of the things that attracted me to computers is that they are consistent and make sense.  If it doesn’t do what you think it should do, you can eventually figure out why and it's perfectly rational and reasonable''~\cite{Computerweekly}.

\textit{Characteristics and Psychological Traits}: 
His professional development has been impacted by his symptoms of depression which, from reports, appears to have played some part in him leaving university twice~\cite{TheGuardian2017}. When he was 29 he was diagnosed with Asperger’s syndrome~\cite{TheGuardian2017}. As he stated,``It's a bit morbid to count the number of times you've had suicidal thoughts, but it was getting to be six to 12 times a day at a peak last winter''~\cite{Independent2}. Reportedly, for him hacking was a form of problem-solving exercise which could have an impact and affect change, just like activism. Research has posited that, ``increased risk of committing cyber-dependent crime is associated with higher autistic-like traits''; however, a diagnosis of autism is not necessarily associated with an increased risk of committing such crime~\cite{Payne2019}. 

\textit{Motivation}: 
Regarding motivation, it is useful to consider some of the key quotes related to this Case. In~\cite{TheGuardian2017} for instance, Case 1 is reported as saying that a hacktivist's ideology ``began to shape his philosophy deeply''. It continued, ``I started to see the power of the internet to make good things happen in the world''. Once again we see a potential sense of a push to use skills for a purpose. In addition, in a sense one may note  a tendency for neutralisation in terms of the potential consequences of his actions~\cite{sykes1957techniques}.

\subsection{Case 5} 
Case 5 was a systems administrator and hacker. He was reportedly accused in 2002 of hacking into a large number of military and NASA computers during a period of 13 months~\cite{Wikipedia}. He became famous in the UK after a protracted attempt by the USA government to have him extradited ultimately ended in failure~\cite{BBCNews2012}. 

\textit{Skills}: 
Case 5 got a computer and practised his technical skills from 14 years old~\cite{BBCNews2012}. After he finished school he went on to become a hairdresser. However, reports~\cite{BBCNews2012} note that his friends later persuaded him to study computers. Following this advice, he completed a computing course and subsequently started work as a contractor in computing. He continued his training in programming and it was these programming skills that he is assumed to have later utilised to hack into government computer systems~\cite{BBCNews2012}.

\textit{Characteristics and Psychological Traits}: 
Case 5 was diagnosed with Asperger’s syndrome during his trial~\cite{BBCNews2012}. This diagnosis lends some explanation to his personality. Reports suggest that Case 1 was introverted and hated leaving his flat~\cite{IEEESpectrum,TheTimes}. Like many people with Asperger’s, there is often a development of highly focused interests. His mother described him as, ``obsessive, naïve, intelligent, ... highly introverted, prone to obsessions and meltdowns and fearful of confrontation'' according to one article~\cite{TheTimes}. As covered by~\cite{IEEESpectrum} his diagnosis may explain his behavior which seemed unreasonable to others.

Case 5 did not see himself as a hacker and was acting alone. Obsessed with UFOs since childhood, reports note that he was convinced that the US was suppressing alien technology and evidence of UFOs~\cite{IEEESpectrum}. As he said, ``I'd stopped washing at one point. I wasn't looking after myself. I wasn't eating properly. I was sitting around the house in my dressing gown, doing this all night'' and to continue, ``I almost wanted to be caught, because it was ruining me. I had this classic thing of wanting to be caught so there would be an end to it''~\cite{BBCNews2012}.

Overall, once again there may be a push or entitlement to use skills for an important purpose as seen in other Cases above (with entitlement linked to other psychological factors~\cite{DarkTriad}). Personality traits such as neuroticism, as defined by Eysenck's theory~\cite{Eysenck1964} are associated with traits such as, depression, anxiety, low-self-esteem, shyness, moodiness, and emotionality. Personality traits such as introversion and neuroticism have also been associated with online criminal behavior~\cite{seigfried2017computer}.

\textit{Motivation}:
In terms of the motive, Case 5 may have committed his acts due to his tendency to form obsessions. He was noted to be obsessed with space and UFOs, and as said above became convinced that the American government was hiding their existence. Allegedly therefore, he hacked into USA military and NASA systems ultimately to prove to himself that UFOs existed~\cite{BBCNews2012}. 
He admitted hacking into US computers but says he had been on a "moral crusade" to find classified documents about UFOs~\cite{BBCNews2012}.
Noting his comments: ``I found out that the US military use Windows and having realised this, I assumed it would probably be an easy hack if they hadn't secured it properly''~\cite{BBCNews2012}.

\section{Discussion and Conclusion}
\label{sec:4}
In exploring how cybercrime occurs, a key component is understanding the nature of attacks and the individuals/actors who have conducted them. This chapter advanced the discussion on cybercriminals (online offenders) with reflection on pertinent literature and an analysis of five prominent cases. From this work, we identified a number of key technology skills that individuals attained throughout their lifetimes, especially in younger years (e.g., Cases 3 and 4). This is by no means definitive but does pose some interesting questions regarding pathways to cybercrime; some of which have been explored before~\cite{Europ,NCA2}. 

There were a range of characteristics and psychological traits covered in the cases including boredom and challenges at school, lower social skills, instability in teenage years, and conditions such as Asperger’s syndrome. Some research (e.g.,~\cite{vblue,Payne2019}) has sought to investigate the links between these factors and online offenders, but clearly more is needed to understand the area given the increasing number and variety of online attacks. To consider the motivation of the cases, in a number of situations there is a push or feeling of entitlement present. This is notable for numerous reasons, but one of the most intriguing is the desire to find the truth or to prevent injustice. These motivations -- as studied here -- are quite different to those of several cybercriminal gangs (e.g., those involved in ransomware or fraud) for instance, who are more motivated by finances.

There are various avenues in the area of cybercriminal profiling and understanding where more research is needed. One of the most important of these is a natural extension of this research and involves a critical examination of a larger set of offender cases. In this work, we concentrated on a number of notorious cases to demonstrate what can be done with openly available reports and data. However, other work could engage with individuals firsthand to understand their profiles and experiences. This may be more representative and not limited (or unduly biased) by cases that feature in the media. Embedding cognitive science and technology into these analyses would provide value for researchers from both fields, and contribute significantly to a more nuanced understanding of cybercrime and its prevention.

\section*{Biography}
\label{sec:bio}

\textbf{Maria Bada} is a Lecturer in Psychology at Queen Mary University in London and a RISCS Fellow in cybercrime. Her research focuses on the human aspects of cybercrime and cybersecurity, such as profiling online offenders, studying their psychologies and pathways towards online deviance as well as the ways to combat cybercrime through tools and capacity building. She is a member of the National Risk Assessment (NRA) Behavioural Science Expert Group in the UK, working on the social and psychological impact of cyber-attacks on members of the public. She has a background in cyberpsychology, and she is a member of the British Psychological Society and the National Counselling Society. 

\noindent \textbf{Jason R.C. Nurse} is an Associate Professor in Cyber Security in the School of Computing at the University of Kent, UK and the Institute of Cyber Security for Society (iCSS), UK. He also holds the roles of Visiting Academic at the University of Oxford, UK and Associate Fellow at the Royal United Services Institute for Defence and Security Studies (RUSI). His research interests include security risk management, corporate communications and cyber security, secure and trustworthy Inter-net of Things, insider threat and cybercrime. He has published over 100 peer-reviewed articles in internationally recognized security journals and conferences.

\bibliographystyle{spmpsci}
\bibliography{main}
\end{document}